# The S66 noncovalent interactions benchmark reconsidered using explicitly correlated methods near the basis set limit[1]

*Manoj K. Kesharwani,*[a] *Amir Karton,*[b] *Nitai Sylvetsky,*[a] *and Jan M.L. Martin*[a,*]

(a) Department of Organic Chemistry, Weizmann Institute of Science, 76100 Reḥovot, Israel. Email: gershom@weizmann.ac.il. FAX: +972 8 934 3029

(b) School of Molecular Sciences, The University of Western Australia, Perth, Western Australia 6009, Australia  E-mail: amir.karton@uwa.edu.au

**Abstract**

The S66 benchmark for noncovalent interactions has been re-evaluated using explicitly correlated methods with basis sets near the one-particle basis set limit. It is found that post-MP2 "high-level corrections" are treated adequately well using a combination of CCSD(F12*) with (aug-)cc-pVTZ-F12 basis sets on the one hand, and (T) extrapolated from conventional CCSD(T)/heavy-aug-cc-pV{D,T}Z on the other hand. Implications for earlier benchmarks on the larger S66x8 problem set in particular, and for accurate calculations on noncovalent interactions in general, are discussed. At a slight cost in accuracy, (T) can be considerably accelerated by using sano-V{D,T}Z+

---

[1] Dedicated to Graham S. Chandler on the occasion of his 80th birthday.

basis sets, while half-counterpoise CCSD(F12*)(T)/cc-pVDZ-F12 offers the best compromise between accuracy and computational cost.

**Introduction**

Noncovalent interactions (NCIs) play crucial roles in supramolecular chemistry and molecular biology, for they are critical in maintaining the three-dimensional structure of large molecules, such as proteins and nucleic acids, and involved in many biological processes in which large molecules bind specifically but temporarily to one another.[1,2]

Significant effort has been devoted to the study and benchmarking of various noncovalently-bonded systems using experimental as well as computational methods (see, e.g., refs[3-10] for recent reviews). However, experimental data are not available in sufficient quantity or in isolation from environmental or dynamical effects, and thus cannot easily be used for the parametrization of approximate methods such as molecular mechanics force fields or semiempirical methods.[11-13] For this reason, wavefunction *ab initio* calculations represent a viable alternative for obtaining highly accurate NCIs.

Where it comes to bond dissociation energies or reaction barrier heights, "chemical accuracy" is typically defined somewhat arbitrarily as ±1 kcal/mol, while the goal of "benchmark accuracy" is more ambitiously defined at ±0.24 kcal/mol (1 kJ/mol).[14] These correspond to relative errors of about 1% or less, as typical bond dissociation energies are on the order of $10^2$ kcal/mol. Noncovalent interactions, on the other hand, are on a much smaller energy scale (an average of 5.5 kcal/mol for the dimers considered in the present study), and a relative accuracy of 1% thus corresponds to about ±0.05 kcal/mol. Such levels of accuracy are, at present, just barely within the reach of wavefunction ab initio methods.[15-18]

Over the past decades, density functional theory (DFT) has become the most widely used electronic structure method in computational quantum chemistry due to its attractive accuracy-to-computational cost ratio. It is well established that the performance of DFT can vary for different types of chemical transformations: generally speaking, the accuracy of a given exchange-correlation (XC) functional should increase as larger molecular fragments are conserved on the two sides of the reaction due to an increasing degree of error cancelation between reactants and products. (For an important caveat, see Ref.[19]) This means that the performance of DFT is better for chemical transformations in which non-covalent interactions are disrupted, compared to transformations in which covalent bonds are broken. That being said, even the best density functional methods are still an order of magnitude less accurate than what is achievable through wavefunction ab initio calculations: two massive survey studies[20,21] on the performance of DFT methods have appeared very recently. As semilocal DFT correlation functionals are intrinsically 'near-sighted', dispersion (which is intrinsically a long-range effect[22]) typically requires either a long-range dispersion correction (see Refs.[23,24] for a review) or a fifth-rung[25] functional[26] such as DSD-PBEP86[27,28] or dRPA75.[29,30] All the most successful functionals for noncovalent interactions require parametrization of either the dispersion correction, or the underlying functional, or both: as a rule of thumb, we advocate that parametrization or validation data should be about an order of magnitude more accurate than the method being parametrized or validated, lest one be merely 'fitting to noise'.

It is well known (e.g., [31]) that second order Møller–Plesset perturbation theory (MP2) is an adequate starting approximation for NCIs (the same cannot be said, however, for molecular atomization energies or reaction barrier heights). Hence, a consensus strategy has emerged in which MP2 interaction energies, obtained using relatively large basis sets, are combined with high level corrections [HLCs = CCSD(T) – MP2] calculated using smaller sets.

However, for interaction energies between bio-molecules and additional realistic-sized systems of interest, the computational cost of HLCs becomes prohibitive even when relatively modest basis sets are used. Therefore, many studies have been devoted to assessing the performance of less-costly density functional methods (with and without dispersion corrections) and low-cost wavefunction methods for standardized NCI benchmarks.

Two popular, and interrelated, such benchmarks are the S66 and S66x8 datasets developed by Hobza and coworkers.[32,33] Both datasets are based around 66 noncovalent dimers, generated from combinations of 14 different monomers, which had been selected based on their frequency as motifs or functional groups in the most commonly found biomolecules. These complexes participate in a wide variety of NCIs, including electrostatic dominated interactions (hydrogen bonding), dispersion dominated interactions (π stacking, aromatic-aliphatic interactions, and aliphatic-aliphatic interactions), and mixed-influence interactions, and are therefore representative of NCIs one might see in biomolecules.

The reference geometries for the S66x8 dataset were obtained by first optimizing each dimer structure at the RI-MP2/cc-pVTZ level, [32,33] then multiplying the intermonomer distances by factors of {0.9,0.95,1.0,1.05,1.10,1.25,1.50,2.00} while keeping the intramonomer geometries frozen—thus generating 8-point unrelaxed "dissociation curves" for each of the 66 monomers. The original S66x8 reference data were computed at the MP2/haV{T,Q}Z level plus a HLC correction computed at the CCSD(T)/AVDZ level, with full counterpoise correction. Hobza and coworkers then carried out quartic interpolation to the {0.9,0.95,1.0,1.05,1.10} points of each curve and determined the minimum of each polynomial, then used those as the reference geometries for the S66 set. That is to say, while the intramolecular geometry in S66 is still MP2/haVTZ level, the intermolecular part is approximately CCSD(T)/CBS. While for many systems (e.g., $(H_2O)_2$, system 1) the difference between S66 and S66x8@1.0$r_e$ is very small, there are quite substantial differences for the π stacking systems such as benzene parallel-displaced dimer (system 24) and stacked uracil dimer (system 26). [These reference geometry differences can give rise to confusion during benchmark calculations if one is not careful.]

Quite recently, the reference values for S66x8 were revised[17] by our group using explicitly correlated MP2 and coupled cluster methods; the revised data were then used for a comprehensive evaluation of many conventional and double-hybrid density functionals.[17] However, due to the relatively-large size of some of the systems in the above datasets (e.g., the uracil dimer), we were unable to obtain HLCs using basis set larger than cc-pVDZ-F12 for the whole set of 528 dimer structures: this might actually be considered as the Achilles' Heel of our study.

Meanwhile, we published a family of diffuse-function augmented basis sets for explicitly correlated calculations, aug-cc-pVnZ-F12 (in short aVnZ-F12).[34] Those were originally developed with anionic systems in mind, but turned out to be beneficial for noncovalent interactions as well, particularly hydrogen bonds.

Now, we were finally able to perform for the whole S66 dataset, at great computational expense, full CCSD(T)-F12b and CCSD(T)(F12*) calculations, as well as CCSD(F12*)/aug-cc-pVTZ and conventional CCSD(T)/heavy-aug-cc-pVTZ (haVTZ for short) calculations. In addition, we were able to treat a subset of 18 systems with still larger cc-pVQZ-F12 and haVQZ basis sets. For both sets of calculations, appropriate counterpoise corrections were also obtained. Thus, firm and robust HLCs for this dataset, that allow assessing the performance of lower level HLCs, are finally within our reach; results and conclusions for such an assessment will therefore be reported in the present work.

**Computational Details**

All calculations at the Weizmann Institute of Science were carried out on the Faculty of Chemistry's Linux cluster "chemfarm", while those at UWA were carried out on the Linux cluster of the Karton group and at the National Computational Infrastructure (NCI) National Facility. Most wavefunction-based *ab initio* calculations were carried out using MOLPRO 2015.1,[35] while ORCA[36] was used for CCSD(2)$_{F12}$,[37] calculations and TURBOMOLE[38] for some additional CCSD(F12)[39] and CCSD[F12] [40,41] calculations.

Conventional, orbital-based, *ab initio* calculations were performed using correlation-consistent[42–45] basis sets. In general, we used the combination of diffuse-function augmented basis sets aug-cc-pVnZ (where n=D,T,Q,5) on nonhydrogen atoms and the regular cc-pVnZ basis sets on hydrogen. For short, this is denoted as haVnZ. In addition, we considered the augmented ano-pVnZ+ and semi-augmented sano-pVnZ+ atomic natural orbital basis sets of Valeev and Neese[46].

For the explicitly correlated MP2-F12, CCSD-F12b,[47,48] and CCSD(F12*)[40,41] (a.k.a., CCSD-F12c) calculations, the correlation consistent cc-pVnZ-F12 basis sets of Peterson et al.[49] were used in conjunction with the appropriate auxiliary basis sets for JKfit[50] (Coulomb and exchange), MP2fit[51,52] (density fitting in MP2), and OptRI [53,54] (complementary auxiliary basis set, CABS) basis sets. For the largest F12-optimized orbital basis set, cc-pV5Z-F12,[55] Weigend's aug-cc-pV5Z/JKFIT basis set[56] for the Coulomb and exchange elements and Hättig's aug-cc-pwCV5Z/MP2FIT basis set[57] for both the RI- MP2 parts and for the CABS; the latter was recommended in Ref. [55] a brute-force alternative, for want of an optimized OptRI. We also employed the aug-cc-pVnZ-F12 basis sets developed in our group[34]; the issue of the appropriate CABS basis set is investigated in detail in Ref.[58]. (See also Ref. [59])

For the purposes of basis set extrapolation, we employed a two-point expression of the form $E(L) = E_\infty + [E(L) - E(L-1)]/\left[\left(L/{L-1}\right)^\alpha - 1\right]$, in which α is taken from Table 2 of Ref[17]. The notation MP2-F12/V{T,Q}Z-F12, for instance, indicates a value extrapolated using this expression with the appropriate α=4.3548 taken from there. For the aV{T,Q}Z-F12 pair, no exponent is given there, and we optimized α=4.6324 by the same procedure as described in Ref.[60]; for aV{D,T}Z-F12, we found α=3.1458.

As suggested in Ref.[60], geminal exponents β=0.9 were set for cc-pVDZ-F12 basis set, β=1.0 were set for cc-pVTZ-F12 and cc-pVQZ-F12 basis sets; for cc-pV5Z-F12, as specified in Ref. [55], β=1.2 was used. CABS correction was used to improve the SCF component.[47,61] For aug-cc-pVnZ-F12, following Ref.[34], we used the same geminal exponents as for the underlying cc-pVnZ-F12 basis sets.

Similar to our previous work[17] three different corrections were considered for the (T) term obtained using explicitly correlated methods:

(a) CCSD(T*)-F12b: the Marchetti-Werner approximation,[62,63] in which the (T) contribution is scaled by the $E_{corr}$[MP2-F12]/$E_{corr}$[MP2] correlation energy ratio.

(b) CCSD(T(b/c))-F12(b/c): (T) is scaled by the respective $E_{corr}$[CCSD-F12(b/c)]/$E_{corr}$[CCSD] ratios.

(c) CCSD(Ts)-F12b:[55] (T) contributions are multiplied by constant scaling factors of 1.1413, 1.0527, and 1.0232 for cc-pVDZ-F12, cc-pVTZ-F12, and cc-pVQZ-F12, respectively (Table 3 in Ref. [55]).

Options (a) and (b) are not strictly size-consistent, but can be rendered so by applying the dimer $E_{corr}$[MP2-F12]/$E_{corr}$[MP2], $E_{corr}$[CCSD-F12(b/c)]/$E_{corr}$[CCSD], ratios also to the monomers: This is indicated by the notation CCSD(T*$_{sc}$) and CCSD(T(b/c)$_{sc}$), respectively.

The treatment of basis set superposition error (BSSE), and in particular the balance between the countervailing forces of BSSE and intrinsic basis set insufficiency (IBSI), has been discussed in Ref.[64] and references therein. Unless basis sets are very close to the 1-particle basis set limit, half-counterpoise (i.e. the average of raw and counterpoise-corrected interaction energies) has been found to yield fastest basis set convergence for both conventional and explicitly correlated interaction energy calculations.[64,7]

Finally, reference geometries were downloaded from BEGDB (http://www.begdb.com)[65] and used verbatim.

**Results and Discussion**

MP2-F12 limit

We were able to perform RI-MP2-F12 calculations with cc-pV5Z-F12 basis set for the entire S66 set, both with and without counterpoise corrections. Unfortunately, the very large auxiliary basis sets that these calculations entail cause a large number of numerical problems due to near-linear dependence: with the aug-cc-pwCV5Z/OptRI CABS basis set, one data point (47, T-shaped benzene dimer) even yielded a plainly absurd result. With the aug-cc-pVQZ/OptRI CABS basis set, we were able to obtain a complete set of data: while of course this CABS basis set does not match the orbital basis set, we recently found[58] (for the S66 dataset at the MP2-F12 level) that CABS basis sets are fairly transferable between similar-sized orbital basis sets.

With the aug-cc-pVQZ-F12 basis set, no numerical problems were encountered except for the need[34] to delete the diffuse f function from carbon atoms to avoid near-linear dependence in the benzene-containing systems. The RMS counterpoise correction over the S66 set can be taken as a gauge for remaining basis set incompleteness: for aVQZ-F12, that amounts to 0.014 kcal/mol at the MP2-F12 level, of which 0.011 kcal/mol comes from the correlation contribution. This is considerably better than 0.042 and 0.034 kcal/mol, respectively, with the cc-pVQZ-F12 basis set, and actually similar to 0.012 kcal/mol (both criteria) obtained with the larger cc-pV5Z-F12 basis set.

Another option is cc-pV{T,Q}Z-F12 basis set extrapolation. For this, the RMS difference between raw and counterpoise-corrected extrapolated values is 0.022 kcal/mol: this drops to 0.006 kcal/mol for cc-pV{Q,5}Z-F12, with the caveat that cc-pV5Z-F12 was obtained using a aug-cc-pVQZ/OptRI CABS.

At any rate, the MP2-F12 component is clearly not the accuracy-limiting factor: indeed, Hobza's[33] best estimated counterpoise corrected MP2 limits differ from counterpoise-corrected RI-MP2-F12/cc-pVQZ-F12 by just 0.009 kcal/mol RMS, which increases to 0.017 kcal/mol relative to aV{T,Q}Z-F12 extrapolation (Table 1).

HLC part 1: CCSD–MP2 difference

The largest basis set for which we were able to perform CCSD(F12*) calculations for the whole set turned out to be aug-cc-pVTZ-F12. The RMS counterpoise correction for that, again used as a gauge for basis set incompleteness, was found to be 0.013 kcal/mol, comparable to the remaining amount for MP2-F12/aug-cc-pVQZ-F12. Intriguingly, the regular cc-pVTZ-F12 yields a functionally equivalent RMS CP of 0.012 kcal/mol: the improvement from the diffuse functions seems to play out primarily at the MP2-F12 level.

For a subset of 18 systems, we were able to perform CCSD(F12*)/cc-pVQZ-F12 calculations, and achieved a reduction of the RMS CP to 0.008 kcal/mol. This improvement is clearly not commensurate with the immensely greater computational expense (about an order of magnitude more CPU time, aside from much greater memory and I/O requirements). This holds especially true in view of the residual uncertainty in the MP2 part.

The CCSD–MP2 difference appears to have stabilized at the CCSD(F12*)/cc-pVQZ-F12 level: the difference between raw and counterpoise-corrected values is just 0.008 kcal/mol RMSD. As justified at great length in Ref.[64] for explicitly correlated calculations with medium and larger basis sets, and in Ref.[7] for conventional calculations with sufficiently large basis sets, we have chosen the half-counterpoise values as our 'gold standard' reference — by construction, these are equidistant from raw and counterpoise-corrected cc-pVQZ-F12 values, by 0.004 kcal/mol RMS (Table 2).

Obviously, cc-pVQZ-F12 is not a realistic option for the entire S66 set. CCSD-F12b/cc-pVTZ-F12 clocks in at 0.015 kcal/mol raw or half-half, and 0.017 kcal/mol with full counterpoise. That drops insignificantly to 0.010 kcal/mol for CCSD(F12*)/cc-pVTZ-F12. With the small cc-pVDZ-F12 basis set, however, CCSD(F12*) has a definite edge over CCSD-F12b, the RMSDs being 0.014 and 0.043 kcal/mol, respectively. This is consistent with what we found in Ref.[17] and applied there for the revised S66x8 dataset. At the CCSD(F12*)/cc-pVDZ-F12 level, raw results have marginally smaller RMSD than half-counterpoise (0.015) and more noticeably smaller than full counterpoise (0.021 kcal/mol;). This is consistent with our findings in Ref.[64], where for small basis sets, uncorrected results consistently agreed better with the basis set limit than counterpoise-corrected ones. (For intermediate basis sets, half-counterpoise works best, while for large basis sets, the choice is largely immaterial as the BSSE corrections are so small.)

The CP-corrected [CCSD-MP2]/AVDZ used in the S66x8 paper[32] here has an RMSD of 0.096 kcal/mol, compared to 0.066 kcal/mol for the CP-corrected [CCSD-MP2]/haV{D,T}Z used in the S66 paper, and just 0.018 kcal/mol for the corresponding *raw* values.

What about the various approximations to CCSD-F12? For the cc-pVDZ-F12 basis set, we were able to obtain full CCSD(F12) values using TURBOMOLE for 64 out of 66 systems (the two uracil dimers proved too large to converge). The RMSD from these values at the CCSD(F12*) level is just 0.001 (!) kcal/mol, making it clear that CCSD(F12*) is, at least for this type of problem, as good an approximation as we can hope for. At the CCSD-F12b level, we incur an error of 0.039 kcal/mol RMS, which increases to 0.067 kcal/mol at the CCSD-F12a level (Table S1, ESI). (The often-proffered claims, e.g. Ref.[66], that CCSD-F12a is superior to CCSD-F12b for small basis sets rest on an error compensation between basis set incompleteness and CCSD-F12a's tendency to overbind.[16]) The CCSD[F12] level, which includes all third-order cross terms from CCSD(F12) but omits the fourth-order terms, clocks in at 0.027 kcal/mol. Using the CCSD(2)$_{F12}$ implementation in ORCA 4, we obtain 0.022 kcal/mol.

It had earlier been suggested to us by reviewers of Refs.[55,16] that the gap between CCSD-F12b and CCSD(F12*) might be closed by evaluating the CABS terms in the projector, which occurs in the dominant CCSD-F12 coupling terms and are neglected in standard CCSD-F12b. (Their evaluation can be forced by setting IXPROJ=1 in MOLPRO.) With the larger cc-pVTZ-F12 basis set, F12b and F12b(IXPROJ=1) are both about 0.02 kcal/mol RMS from CCSD(F12*): In the cc-pVDZ-F12 basis set, however, a larger benefit is observed: RMSD=0.018 kcal/mol from full CCSD(F12) is actually better than CCSD[F12] (see above). We noted previously for the water clusters[16] that the projector terms do converge very rapidly with the basis set.

HLC part 2: triples (T) term

This term does not benefit from F12, although various approximate scaling techniques have been proposed (see below).

It was previously found for the water clusters[16] as well as for main-group atomization energies[67] that (T) is best obtained from conventional CCSD(T) calculations. We will attempt to do so here.

For a subset of 18 systems, we were able to do CCSD(T)/haV{T,Q}Z extrapolation: the RMS counterpoise correction to (T) is just 0.004 kcal/mol: This suggests the extrapolated value is very close to the 1-particle infinite basis set limit, where said difference should vanish. It also means that these values should be an acceptable benchmark for lower-level approaches.

Intriguingly, for CCSD(T)/sano-V{T,Q}Z+ the RMS CP correction increases to 0.010 kcal/mol, even though atomic natural orbital basis sets should (in principle) minimize basis set superposition error. Extrapolation of raw and counterpoise-corrected (T)/haV{T,Q}Z contributions to the interaction energy led to nearly identical results, the RMS difference between both sets of results being a negligible 0.007 kcal/mol (Table 3). This suggests the extrapolated value is very close to the 1-particle infinite basis set limit, where said difference should vanish. We have somewhat arbitrarily chosen the average of both values — mathematically equivalent to "half-counterpoise" — as our "gold standard" reference value.

The (T)/haV{D,T}Z does remarkably well, at an RMSD relative to half-CP (T)/haV{T,Q}Z of just 0.011 kcal/mol without counterpoise and 0.008 kcal/mol with half counterpoise. The (T) for uracil dimer still takes over a week wall time on 16 cores, though: with the smaller sano-V{D,T}Z+ basis sets, the calculation time can be halved, at the expense of increasing RMSD to 0.022 kcal/mol (raw) and 0.027 kcal/mol (half-CP). Thus, sano-V{D,T}Z may be a viable option where haV{D,T}Z is computationally too expensive and various F12 scaling schemes with (aug-)cc-pVDZ-F12 basis sets too inaccurate.

What about the various scaling schemes for the triples? With the cc-pVTZ-F12 basis set, which is similar in cost, RMSDs over the small 18-system set are smallest for (Ts) with half-counterpoise (0.006 kcal/mol), (T*$_{sc}$) with full counterpoise (0.007 kcal/mol), or unscaled (T) without counterpoise correction (0.012 kcal/mol, similar to (Tb$_{sc}$) half-counterpoise) (Table 4). For VQZ-F12, very low RMSDs are obtained for (Ts) with half-counterpoise, (Tb$_{sc}$) or (T*$_{sc}$) with full counterpoise, or unscaled raw (T).

We now turn to the cc-pVDZ-F12 basis set, where recovery of the (T) term represents a challenge on account of the small basis set. With either the CCSD(F12*) or the CCSD-F12b ansatz, half-counterpoise is clearly superior over the two other choices, particularly CCSD(Tc$_{sc}$) and CCSD(T*$_{sc}$).

### HLC considered as a whole

The best level of HLC we can afford, and that only for a subset of 18 systems, would be what we could term GOLD: [CCSD(F12*)–MP2-F12]/cc-pVQZ-F12 half-counterpoise combined with (T)/haV{T,Q}Z. The next level down we term SILVER: CCSD(F12*)/aug-cc-pVTZ-F12 half-counterpoise combined with (T)/haV{D,T}Z , which we were able to complete for all 66 systems. (By far the most CPU-intensive step was the CCSD(T)/haVTZ calculations, which took over 2 weeks on a 16-CPU machine for uracil dimer.)

The RMS difference between GOLD and SILVER for the 18 systems where we have the former available is just 0.006 kcal/mol, hence we are probably justified using SILVER as a standard. For comparison, full-CP [CCSD(T)–MP2]/haV{D,T}Z as used for the Hobza S66 reference values has an RMSD=0.053 kcal/mol from SILVER, and 0.054 kcal/mol from GOLD.

As unlike for the MP2-F12 part, there seems to be comparatively little effect of the extra diffuse functions on the [CCSD(F12*)–MP2-F12] part, we also considered a lower-cost STERLING level (i.e., silver-copper alloy) in which [CCSD(F12*)–MP2-F12]/aug-cc-pVTZ-F12 is combined with (T)/sano-V{D,T}Z+. Deviation from GOLD is just 0.022 kcal/mol RMS over the 18-system subset, from SILVER it is 0.026 kcal/mol over the whole S66 set. As the most expensive calculation step for the largest systems in S66 has now been reduced to "just" five days on a 16-core machine, this may be a viable option for larger benchmarks if greater accuracy is needed.

Can we avoid having to do the (T) in a triple-zeta basis altogether? This pretty much implies using some form of scaling scheme. From Table 5, it would seem that CCSD(F12*)(Tc$_{sc}$)/cc-pVDZ-F12 with half-counterpoise would be the lowest-cost option. According to Table 5, this deviates by just 0.032 kcal/mol RMS from SILVER, and hence becomes our new BRONZE (or BRONZEnew) option.

The same level of HLC without counterpoise was used in the S66x8 revision paper, and alas incurs RMS=0.096 kcal/mol, even though the deviation from GOLD for the 18-system subset is just 0.044 kcal/mol. The difference between raw CCSD(F12*)(Tc$_{sc}$)/cc-pVDZ-F12 and its half-CP counterpart can indeed be quite nontrivial, reaching a maximum of 0.2 kcal/mol for system 26 (stacked uracil dimer).

Final reference data and overall performance

Our final reference data are given in Table 6, while Figure 1 graphically represents the errors for the entire S66 set compared to SILVER references. The Hobza S66 reference values are actually quite close to SILVER : unfortunately, the CCSD(T)/haVTZ step they entail makes them essentially as expensive as SILVER itself. Aside from a few outliers (acetic acid and acetamide dimers), STERLING performs quite well and might be a viable option for a recalculation of S66x8. BRONZEnew would be the next best solution, but is computationally much less expensive: the extra cost compared to our published S66x8 revision is essentially that of the CCSD(F12*)(T)/cc-pVDZ-F12 counterpoise steps. Considering that this cuts the interquartile range of the errors approximately in half, we believe that the fairly modest extra cost is well justified.

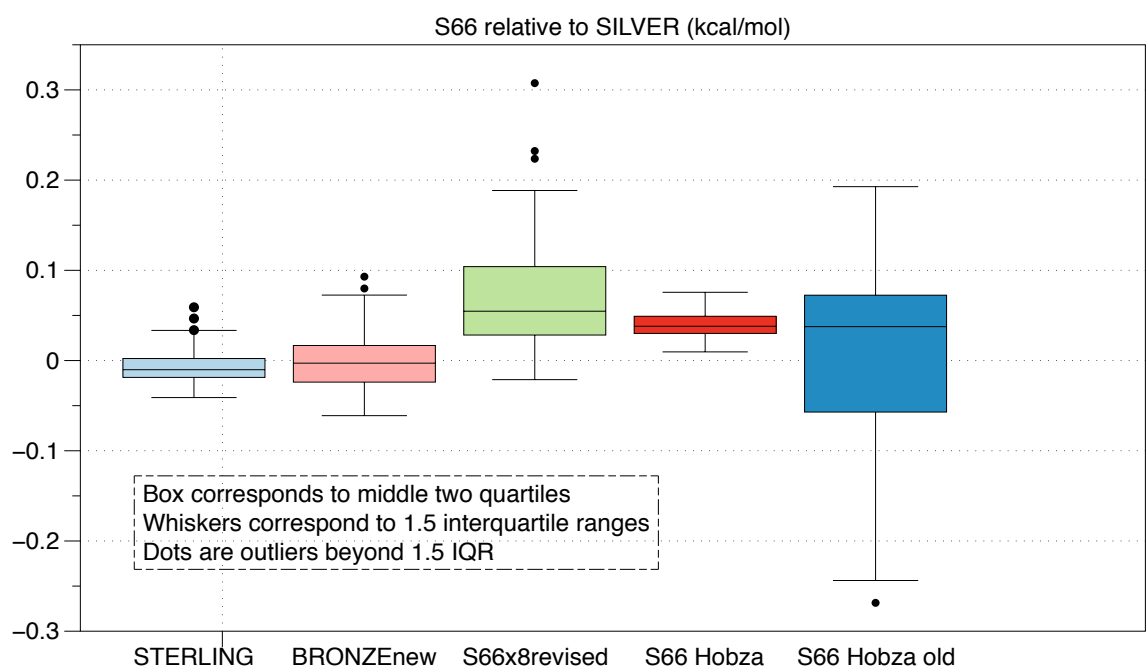

**Figure 1: box plot of errors in S66 association energies compared to SILVER values (kcal/mol)**

Finally, the "old S66" level of theory, at which Hobza and coworkers calculated the S66 set, clearly has an inadequate HLC (just CCSD(T)/AVDZ with full counterpoise), as noted previously in Ref.[17]

CP-corrected [CCSD(T)-MP2]/AVDZ, as used in the S66x8 paper, clearly does benefit from some error compensation between CCSD–MP2 and (T), as the RMSD relative to the 66-system silver standard is *substantially* smaller than the RMSDs on the two constituent components.

A preliminary update of S66x8

It is computationally quite feasible to re-evaluate all of S66x8 at our new BRONZE level. These data are presented in Table 7.

Differences with our previously published revision are small at equilibrium and stretched distances (0.053 kcal/mol RMS), but more significant at compressed distances (0.108 kcal/mol RMS). The largest differences are seen for dimers involving uracil. We attempted aug-cc-pVDZ-F12 half-CP as well, but spot-checking against SILVER results for a small subset of systems (20, 21, 24) suggests that any calculation in which (T) is not evaluated with at least a haVTZ basis set needs to be considered with caution for the compressed distances. Such a re-evaluation is presently in progress, but will take time owing to its formidable computational cost for particularly the uracil dimers.

**Conclusions**

We have re-evaluated the S66 benchmark for noncovalent interactions at the highest ab initio level that is currently feasible. Obtaining reliable MP2-F12 basis set limits does not appear to represent a serious challenge: our aug-cc-pV{T,Q}Z-F12 extrapolated values are probably as reliable as can be achieved, and are only a minor factor in the cost of the overall benchmark calculations.

We can define three tiers of accuracy for the CCSD(T)-MP2 "high-level correction" HLC:

- "GOLD" combining [CCSD(F12*)–MP2-F12]/cc-pVQZ-F12 half-CP with (T)/haV{T,Q}Z half-CP
- "SILVER" combining [CCSD(F12*)–MP2-F12]/aug-cc-pVTZ-F12 half-CP with (T)/haV{D,T}Z half-CP
- "BRONZE", i.e., half-counterpoise CCSD(F12*)(Tc$_{sc}$)/cc-pVDZ-F12

In addition, we can identify a reduced-cost variant of SILVER, which we will call

- "STERLING" combining [CCSD(F12*)–MP2-F12]/cc-pVTZ-F12 half-CP with (T)/sano-PV{D,T}Z+ raw

SILVER is available for the entire set: for a subset of 18 systems where GOLD was feasible, SILVER deviates from it by less than 0.01 kcal/mol RMS.

STERLING sacrifices about 0.02 kcal/mol RMS in accuracy, but cuts overall computation time in half thanks to the smaller 'semi-augmented' basis sets used for the (T).

BRONZE deviates by just over 0.03 kcal/mol RMS from SILVER, but at drastically reduced cost (over an order of magnitude) and is a viable option for larger systems and more extensive benchmarks.

The revised S66 benchmark of the Hobza group stands up well under scrutiny: the fairly inexpensive CCSD(T)/AVDZ HLCs used in their S66x8 study are inadequate, but still do benefit from error compensation. The revised S66x8 values from our group could still have been improved further by adding half-counterpoise corrections to the HLC. We have presented such values in the present work; however, we recommend that they be treated with caution, particularly at compressed distances. A more thorough re-evaluation is in progress.

**Acknowledgments**

N.S. acknowledges a graduate fellowship from the Feinberg Graduate School. Research at Weizmann was supported by the Israel Science Foundation (grant 1358/15), by the Minerva Foundation (Munich, Germany), and by the Helen and Martin Kimmel Center for Molecular Design (Weizmann Institute of Science).

Research at UWA was undertaken with the assistance of resources from the National Computational Infrastructure (NCI), which is supported by the Australian Government. We also acknowledge the system administration support provided by the Faculty of Science at the University of Western Australia to the Linux cluster of the Karton group. A.K. acknowledges the Australian Research Council for a Future Fellowship (FT170100373).

**Conflict of interest statement**

The authors declare no conflict of interest.

**Supporting information**

Excel spreadsheet containing calculated total energies and interaction energies for the S66 dataset at the levels of theory considered.

**Table 1:** RMSD (kcal/mol) for the MP2-F12 limits of the S66 set

|  | relative to cc-pV{Q,5}Z-F12 half-CP | | | relative to aV{T,Q}Z-F12 half-CP | | |
| --- | --- | --- | --- | --- | --- | --- |
|  | Raw | CP | Half | Raw | CP | Half |
| cc-pVDZ-F12 | 0.083 | 0.148 | 0.042 | 0.085 | 0.145 | 0.040 |
| cc-pVTZ-F12 | 0.066 | 0.052 | 0.014 | 0.068 | 0.050 | 0.015 |
| cc-pVQZ-F12 | 0.029 | 0.014 | 0.009 | 0.032 | 0.011 | 0.011 |
| cc-pV5Z-F12 | 0.007 | 0.006 | 0.002 |  |  |  |
| cc-pV{T,Q}Z-F12 | 0.023 | 0.005 | 0.012 | 0.025 | 0.006 | 0.015 |
| cc-pV{Q,5}Z-F12 | 0.003 | 0.003 | REF |  |  |  |
| aVDZ-F12 | 0.088 | 0.092 | 0.030 | 0.091 | 0.090 | 0.031 |
| aVTZ-F12 | 0.038 | 0.019 | 0.011 | 0.040 | 0.017 | 0.014 |
| aVQZ-F12 | 0.008 | 0.007 | 0.003 | 0.011 | 0.005 | 0.004 |
| aV{D,T}Z-F12 | 0.040 | 0.009 | 0.024 | 0.043 | 0.011 | 0.027 |
| aV{T,Q}Z-F12 | 0.007 | 0.003 | 0.004 | 0.004 | 0.004 | REF |
| MP2/CBS[a] |  | 0.019 |  |  | 0.017 |  |

[a]MP2 basis set limit used by the Hobza and coworkers for S66 dataset

**Table 2:** RMS Deviations (kcal/mol) for the CCSD–MP2 components from the basis set limit values of S66 interaction energies as calculated with various basis sets

|  | relative to GOLD[a] reference (for 18 systems) | | | relative to SILVER[b] reference (for complete S66 set) | | |
|---|---|---|---|---|---|---|
|  | Raw | CP | Half | Raw | CP | Half |
| **CCSD-F12b–MP2-F12** | | | | | | |
| cc-pVDZ-F12 | 0.043 | 0.063 | 0.052 | 0.042 | 0.051 | 0.044 |
| cc-pVTZ-F12 | 0.015 | 0.017 | 0.015 | 0.019 | 0.017 | 0.017 |
| cc-pVQZ-F12 | 0.005 | 0.006 | 0.004 | | | |
| **CCSD-(F12*)–MP2-F12** | | | | | | |
| cc-pVDZ-F12 | 0.014 | 0.021 | 0.015 | 0.015 | 0.020 | 0.014 |
| cc-pVTZ-F12 | 0.010 | 0.007 | 0.006 | 0.011 | 0.007 | 0.007 |
| cc-pVQZ-F12 | 0.004 | 0.004 | REF | | | |
| aVDZ-F12 | 0.016 | 0.015 | 0.006 | 0.029 | 0.020 | 0.009 |
| aVTZ-F12 | 0.010 | 0.005 | 0.007 | 0.006 | 0.006 | REF |
| **CCSD–MP2** | | | | | | |
| haVDZ | 0.074 | 0.098 | 0.080 | 0.093 | 0.183 | 0.133 |
| haVTZ | 0.018 | 0.063 | 0.036 | 0.020 | 0.088 | 0.044 |
| haV{D,T}Z | 0.018 | 0.066 | 0.039 | 0.031 | 0.060 | 0.025 |
| haVQZ | 0.014 | 0.053 | 0.022 | | | |
| sano-pVDZ+ | 0.138 | 0.145 | 0.137 | 0.210 | 0.267 | 0.237 |
| sano-pVTZ+ | 0.085 | 0.067 | 0.055 | 0.071 | 0.113 | 0.074 |
| sano-pVQZ+ | 0.048 | 0.057 | 0.022 | | | |
| AVDZ[c] | | 0.096 | | | 0.179 | |

[a] GOLD: CCSD(F12*)–MP2-F12/cc-pVQZ-F12 half-CP (for subset of 18 out of 66 systems)
[b] SILVER: CCSD(F12*)–MP2-F12/aVTZ-F12 half-CP (for complete S66 set)
[c] Level of theory used by the Hobza and coworkers in original S66 dataset

**Table 3:** RMS Deviations (kcal/mol) for the (T) term of conventional CCSD(T) calculated for S66 interaction energies with various basis sets.

|  | For 18 sub-systems | | | For complete S66 set | | |
|---|---|---|---|---|---|---|
|  | Raw | CP | Half | Raw | CP | Half |
| haV{T,Q}Z | 0.002 | 0.002 | REF[a] |  |  |  |
| haV{D,T}Z | 0.007 | 0.011 | 0.008 | 0.007 | 0.007 | REF[b] |
| haVQZ | 0.005 | 0.019 | 0.012 |  |  |  |
| haVTZ | 0.011 | 0.047 | 0.028 | 0.022 | 0.052 | 0.026 |
| haVDZ | 0.035 | 0.137 | 0.081 | 0.086 | 0.172 | 0.089 |
| sano-pV{T,Q}Z+ | 0.010 | 0.007 | 0.007 |  |  |  |
| sano-pV{D,T}Z+ | 0.022 | 0.041 | 0.027 | 0.026 | 0.057 | 0.039 |
| sano-pVQZ+ | 0.020 | 0.043 | 0.030 |  |  |  |
| sano-pVTZ+ | 0.047 | 0.095 | 0.069 | 0.073 | 0.132 | 0.102 |
| sano-pVDZ+ | 0.128 | 0.226 | 0.176 | 0.199 | 0.312 | 0.254 |
| AVDZ[c] |  | 0.115 |  |  | 0.134 |  |

[a] GOLD: [CCSD(T)−CCSD]/haV{T,Q}Z half-CP (for subset of 18 out of 66 systems)
[b] SILVER: [CCSD(T)−CCSD]/haV{D,T}Z half-CP reference (for complete S66 set)
[c] Level of theory used by the Hobza and coworkers in original S66 dataset

**Table 4:** RMS Deviations (kcal/mol) for the (T) term of explicitly correlated CCSD(T)-F12x calculated for S66 interaction energies with various basis sets.

|  | Raw (T*) | Raw (T) | Raw (Tb$_{sc}$) | Raw (T*$_{sc}$) | Raw (Ts) | CP (T*) | CP (T) | CP (Tb$_{sc}$) | CP (T*$_{sc}$) | CP (Ts) | Half (T*) | Half (T) | Half (Tb$_{sc}$) | Half (T*$_{sc}$) | Half (Ts) |
|---|---|---|---|---|---|---|---|---|---|---|---|---|---|---|---|
| **relative to GOLD[a] reference (for 18 sub-systems)** | | | | | | | | | | | | | | | |
| F12b/cc-pVQZ-F12 | 0.022 | 0.006 | 0.020 | 0.025 | 0.016 | 0.010 | 0.019 | 0.003 | 0.003 | 0.007 | 0.015 | 0.008 | 0.009 | 0.013 | 0.005 |
| F12b/cc-pVTZ-F12 | 0.042 | 0.012 | 0.037 | 0.045 | 0.030 | 0.017 | 0.046 | 0.015 | 0.007 | 0.021 | 0.027 | 0.024 | 0.012 | 0.020 | 0.006 |
| F12b/cc-pVDZ-F12 | 0.066 | 0.046 | 0.045 | 0.061 | 0.033 | 0.044 | 0.132 | 0.067 | 0.053 | 0.077 | 0.038 | 0.088 | 0.018 | 0.013 | 0.027 |
| (F12*)/cc-pVDZ-F12 | 0.057 | 0.055 | 0.039 | 0.050 | 0.025 | 0.051 | 0.141 | 0.074 | 0.063 | 0.086 | 0.036 | 0.097 | 0.024 | 0.015 | 0.037 |
| (F12*)/aVDZ-F12 | 0.055 | 0.043 | 0.045 | 0.058 | 0.032 | 0.035 | 0.117 | 0.048 | 0.037 | 0.060 | 0.036 | 0.079 | 0.013 | 0.017 | 0.020 |
| **relative to SILVER[b] (for complete S66 set)** | | | | | | | | | | | | | | | |
| F12b/cc-pVTZ-F12 | 0.062 | 0.018 | 0.066 | 0.079 | 0.057 | 0.019 | 0.054 | 0.010 | 0.012 | 0.015 | 0.039 | 0.023 | 0.032 | 0.044 | 0.023 |
| F12b/cc-pVDZ-F12 | 0.085 | 0.044 | 0.096 | 0.122 | 0.080 | 0.048 | 0.167 | 0.064 | 0.044 | 0.076 | 0.041 | 0.103 | 0.029 | 0.047 | 0.022 |
| (F12*)/cc-pVDZ-F12 | 0.069 | 0.058 | 0.084 | 0.103 | 0.063 | 0.061 | 0.181 | 0.074 | 0.058 | 0.092 | 0.036 | 0.118 | 0.023 | 0.032 | 0.026 |
| (F12*)/aVDZ-F12 | 0.080 | 0.032 | 0.121 | 0.142 | 0.102 | 0.035 | 0.140 | 0.036 | 0.028 | 0.049 | 0.045 | 0.081 | 0.055 | 0.072 | 0.040 |

[a]GOLD: [CCSD(T)–CCSD]/haV{T,Q}Z half-CP (for subset of 18 out of 66 systems)
[b]SILVER: [CCSD(T)–CCSD]/haV{D,T}Z half-CP reference (for complete S66 set)

**Table 5:** RMS Deviations (kcal/mol) for the high level corrections (HLC = [CCSD(T)-F12x – MP2-F12]/cc-pVnZ-F12)]) components of the S66 interaction energies.

| | Raw HLC (T*) | Raw HLC (T) | Raw HLC (T(b/c)$_{sc}$) | Raw HLC (T*$_{sc}$) | Raw HLC (Ts) | CP HLC (T*) | CP HLC (T) | CP HLC (T(b/c)$_{sc}$) | CP HLC (T*$_{sc}$) | CP HLC (Ts) | Half HLC (T*) | Half HLC (T) | Half HLC (T(b/c)$_{sc}$) | Half HLC (T*$_{sc}$) | Half HLC (Ts) |
|---|---|---|---|---|---|---|---|---|---|---|---|---|---|---|---|
| **relative to GOLD[a] reference (for 18 sub-systems)** | | | | | | | | | | | | | | | |
| F12b/cc-pVQZ-F12 | 0.019 | 0.003 | 0.018 | 0.023 | 0.013 | 0.014 | 0.014 | 0.005 | 0.009 | 0.004 | 0.016 | 0.007 | 0.011 | 0.015 | 0.007 |
| F12b/cc-pVTZ-F12 | 0.039 | 0.009 | 0.035 | 0.043 | 0.028 | 0.020 | 0.055 | 0.026 | 0.021 | 0.032 | 0.023 | 0.029 | 0.012 | 0.018 | 0.012 |
| F12b/cc-pVDZ-F12 | 0.039 | 0.082 | 0.024 | 0.031 | 0.027 | 0.089 | 0.184 | 0.122 | 0.110 | 0.133 | 0.048 | 0.132 | 0.063 | 0.051 | 0.075 |
| (F12*)/cc-pVDZ-F12 | 0.057 | 0.058 | 0.043 | 0.053 | 0.032 | 0.061 | 0.153 | 0.087 | 0.077 | 0.100 | 0.036 | 0.103 | 0.034 | 0.026 | 0.046 |
| (F12*)/aVDZ-F12 | 0.067 | 0.031 | 0.060 | 0.073 | 0.047 | 0.042 | 0.130 | 0.062 | 0.052 | 0.074 | 0.036 | 0.079 | 0.016 | 0.020 | 0.023 |
| **relative to SILVER[b] (for complete S66 set)** | | | | | | | | | | | | | | | |
| F12b/cc-pVTZ-F12 | 0.068 | 0.025 | 0.075 | 0.088 | 0.066 | 0.021 | 0.057 | 0.022 | 0.024 | 0.026 | 0.041 | 0.024 | 0.039 | 0.051 | 0.032 |
| F12b/cc-pVDZ-F12 | 0.058 | 0.072 | 0.078 | 0.102 | 0.065 | 0.074 | 0.196 | 0.103 | 0.087 | 0.115 | 0.035 | 0.132 | 0.050 | 0.053 | 0.055 |
| (F12*)/cc-pVDZ-F12 | 0.073 | 0.054 | 0.092 | 0.112 | 0.072 | 0.069 | 0.190 | 0.087 | 0.071 | 0.103 | 0.034 | 0.119 | 0.032 | 0.039 | 0.035 |
| (F12*)/aVDZ-F12 | 0.105 | 0.035 | 0.149 | 0.170 | 0.130 | 0.044 | 0.158 | 0.053 | 0.041 | 0.067 | 0.048 | 0.078 | 0.061 | 0.078 | 0.047 |

[a]GOLD: half-counterpoise corrected [CCSD(F12*) –MP2-F12]/cc-pVQZ-F12 combined with half-counterpoise corrected [CCSD(T)–CCSD]/haV{T,Q}Z
[b]SILVER: half-counterpoise corrected [CCSD(F12*)–MP2-F12]/aVTZ-F12 combined with half-counterpoise corrected [CCSD(T)–CCSD]/haV{D,T}Z

**Table 6:** Systems in the S66 dataset and final recommended dissociation energies (kcal/mol) obtained in the present work.

| Systems | GOLD[a] | SILVER[b] | Systems | GOLD[a] | SILVER[b] |
|---|---|---|---|---|---|
| 01 Water ... Water | 4.979 | 4.982 | 34 Pentane ... Pentane | | 3.741 |
| 02 Water ... MeOH | 5.666 | 5.666 | 35 Neopentane ... Pentane | | 2.582 |
| 03 Water ... MeNH$_2$ | 6.985 | 6.986 | 36 Neopentane ... Neopentane | | 1.745 |
| 04 Water ... Peptide | | 8.183 | 37 Cyclopentane ... Neopentane | | 2.376 |
| 05 MeOH ... MeOH | 5.824 | 5.822 | 38 Cyclopentane ... Cyclopentane | | 2.967 |
| 06 MeOH ... MeNH$_2$ | 7.625 | 7.617 | 39 Benzene ... Cyclopentane | | 3.488 |
| 07 MeOH ... Peptide | | 8.307 | 40 Benzene ... Neopentane | | 2.824 |
| 08 MeOH ... Water | 5.065 | 5.064 | 41 Uracil ... Pentane | | 4.761 |
| 09 MeNH$_2$ ... MeOH | 3.088 | 3.087 | 42 Uracil ... Cyclopentane | | 4.052 |
| 10 MeNH$_2$ ... MeNH$_2$ | 4.189 | 4.184 | 43 Uracil ... Neopentane | | 3.652 |
| 11 MeNH$_2$ ... Peptide | | 5.436 | 44 Ethene ... Pentane | | 1.973 |
| 12 MeNH$_2$ ... Water | 7.354 | 7.349 | 45 Ethyne ... Pentane | | 1.696 |
| 13 Peptide ... MeOH | | 6.251 | 46 Peptide ... Pentane | | 4.215 |
| 14 Peptide ... MeNH$_2$ | | 7.516 | 47 Benzene ... Benzene (TS) | | 2.801 |
| 15 Peptide ... Peptide | | 8.689 | 48 Pyridine ... Pyridine (TS) | | 3.472 |
| 16 Peptide ... Water | | 5.180 | 49 Benzene ... Pyridine (TS) | | 3.260 |
| 17 Uracil ... Uracil (BP) | | 17.407 | 50 Benzene ... Ethyne (CH-π) | 2.839 | 2.828 |
| 18 Water ... Pyridine | | 6.927 | 51 Ethyne ... Ethyne (TS) | 1.526 | 1.519 |
| 19 MeOH ... Pyridine | 7.464 | 7.467 | 52 Benzene ... AcOH (OH-π) | | 4.691 |
| 20 AcOH ... AcOH | 19.364 | 19.361 | 53 Benzene ... AcNH$_2$ (NH-π) | | 4.376 |
| 21 AcNH$_2$ ... AcNH$_2$ | 16.468 | 16.474 | 54 Benzene ... Water (OH-π) | | 3.267 |
| 22 AcOH ... Uracil | | 19.736 | 55 Benzene ... MeOH (OH-π) | | 4.139 |
| 23 AcNH$_2$ ... Uracil | | 19.420 | 56 Benzene ... MeNH$_2$ (NH-π) | | 3.174 |
| 24 Benzene ... Benzene (π-π) | | 2.685 | 57 Benzene ... Peptide (NH-π) | | 5.222 |
| 25 Pyridine ... Pyridine (π-π) | | 3.751 | 58 Pyridine ... Pyridine (CH-N) | | 4.189 |
| 26 Uracil ... Uracil (π-π) | | 9.672 | 59 Ethyne ... Water (CH-O) | 2.912 | 2.905 |
| 27 Benzene ... Pyridine (π-π) | | 3.300 | 60 Ethyne ... AcOH (OH-π) | 4.925 | 4.917 |
| 28 Benzene ... Uracil (π-π) | | 5.517 | 61 Pentane ... AcOH | | 2.876 |
| 29 Pyridine ... Uracil (π-π) | | 6.629 | 62 Pentane ... AcNH$_2$ | | 3.491 |
| 30 Benzene ... Ethene | 1.348 | 1.358 | 63 Benzene ... AcOH | | 3.709 |
| 31 Uracil ... Ethene | | 3.291 | 64 Peptide ... Ethene | | 2.967 |
| 32 Uracil ... Ethyne | | 3.651 | 65 Pyridine ... Ethyne | | 4.064 |
| 33 Pyridine ... Ethene | 1.790 | 1.779 | 66 MeNH$_2$ ... Pyridine | | 3.930 |

[a]GOLD: MP2-F12/aV{T,Q}Z-F12 half-CP + [CCSD(F12*) −MP2-F12]/cc-pVQZ-F12 half-CP + [CCSD(T)−CCSD]/haV{T,Q}Z half-CP
[b]SILVER: MP2-F12/aV{T,Q}Z-F12 half-CP + [CCSD(F12*)−MP2-F12]/aVTZ-F12 half-CP + [CCSD(T)−CCSD]/haV{D,T}Z half-CP

**Table 7:** Preliminary re-evaluation of the S66x8 dataset at our new BRONZE level (kcal/mol)

|  | $0.9r_e$ | $0.95r_e$ | $1.0r_e$ | $1.05r_e$ | $1.1r_e$ | $1.25r_e$ | $1.5r_e$ | $2.0r_e$ |
|---|---|---|---|---|---|---|---|---|
| 01 Water ... Water | 4.610 | 4.912 | 4.915 | 4.739 | 4.462 | 3.460 | 2.108 | 0.871 |
| 02 Water ... MeOH | 5.232 | 5.578 | 5.589 | 5.396 | 5.086 | 3.947 | 2.385 | 0.952 |
| 03 Water ... MeNH$_2$ | 6.530 | 6.897 | 6.894 | 6.660 | 6.291 | 4.916 | 2.973 | 1.140 |
| 04 Water ... Peptide | 7.666 | 8.076 | 8.089 | 7.851 | 7.465 | 5.991 | 3.828 | 1.437 |
| 05 MeOH ... MeOH | 5.328 | 5.727 | 5.773 | 5.599 | 5.299 | 4.147 | 2.526 | 1.010 |
| 06 MeOH ... MeNH$_2$ | 7.006 | 7.497 | 7.558 | 7.346 | 6.971 | 5.497 | 3.342 | 1.273 |
| 07 MeOH ... Peptide | 7.689 | 8.186 | 8.255 | 8.051 | 7.680 | 6.190 | 3.649 | 1.099 |
| 08 MeOH ... Water | 4.629 | 4.985 | 5.023 | 4.867 | 4.601 | 3.595 | 2.201 | 0.908 |
| 09 MeNH$_2$ ... MeOH | 2.813 | 3.037 | 3.029 | 2.889 | 2.680 | 1.970 | 1.097 | 0.394 |
| 10 MeNH$_2$ ... MeNH$_2$ | 3.671 | 4.068 | 4.127 | 3.985 | 3.730 | 2.777 | 1.302 | 0.388 |
| 11 MeNH$_2$ ... Peptide | 4.922 | 5.334 | 5.387 | 5.221 | 4.926 | 3.204 | 1.402 | 0.457 |
| 12 MeNH$_2$ ... Water | 6.777 | 7.224 | 7.259 | 7.036 | 6.659 | 5.217 | 3.149 | 1.195 |
| 13 Peptide ... MeOH | 5.725 | 6.149 | 6.213 | 6.053 | 5.762 | 4.615 | 2.952 | 1.308 |
| 14 Peptide ... MeNH$_2$ | 6.860 | 7.379 | 7.478 | 7.310 | 6.978 | 5.611 | 3.552 | 1.492 |
| 15 Peptide ... Peptide | 8.085 | 8.590 | 8.673 | 8.485 | 8.131 | 6.677 | 4.426 | 1.782 |
| 16 Peptide ... Water | 4.740 | 5.097 | 5.146 | 5.006 | 4.758 | 3.805 | 2.458 | 1.136 |
| 17 Uracil ... Uracil (BP) | 16.042 | 17.190 | 17.432 | 17.078 | 16.352 | 13.263 | 8.410 | 3.357 |
| 18 Water ... Pyridine | 6.453 | 6.853 | 6.872 | 6.653 | 6.294 | 4.939 | 3.013 | 1.189 |
| 19 MeOH ... Pyridine | 6.855 | 7.366 | 7.451 | 7.263 | 6.911 | 5.490 | 3.385 | 1.336 |
| 20 AcOH ... AcOH | 17.793 | 19.070 | 19.328 | 18.923 | 18.106 | 14.657 | 9.246 | 3.595 |
| 21 AcNH$_2$ ... AcNH$_2$ | 15.185 | 16.242 | 16.441 | 16.085 | 15.386 | 12.485 | 8.022 | 3.009 |
| 22 AcOH ... Uracil | 18.241 | 19.470 | 19.732 | 19.358 | 18.581 | 15.243 | 9.907 | 4.161 |
| 23 AcNH$_2$ ... Uracil | 18.011 | 19.157 | 19.419 | 19.094 | 18.391 | 15.309 | 10.276 | 4.672 |
| 24 Benzene ... Benzene (π-π) | -0.030 | 1.905 | 2.634 | 2.739 | 2.546 | 1.543 | 0.499 | 0.067 |
| 25 Pyridine ... Pyridine (π-π) | 1.063 | 2.997 | 3.716 | 3.788 | 3.543 | 2.360 | 0.974 | 0.241 |
| 26 Uracil ... Uracil (π-π) | 7.693 | 9.396 | 9.765 | 9.406 | 8.688 | 6.110 | 3.138 | 1.013 |
| 27 Benzene ... Pyridine (π-π) | 0.463 | 2.523 | 3.265 | 3.338 | 3.096 | 1.976 | 0.735 | 0.151 |
| 28 Benzene ... Uracil (π-π) | 3.278 | 5.041 | 5.588 | 5.475 | 5.034 | 3.296 | 1.380 | 0.260 |
| 29 Pyridine ... Uracil (π-π) | 3.460 | 5.990 | 6.698 | 6.513 | 5.936 | 3.876 | 1.796 | 0.543 |
| 30 Benzene ... Ethene | 0.048 | 0.977 | 1.310 | 1.333 | 1.210 | 0.678 | 0.176 | -0.008 |
| 31 Uracil ... Ethene | 2.424 | 3.130 | 3.300 | 3.181 | 2.921 | 1.981 | 0.937 | 0.257 |
| 32 Uracil ... Ethyne | 2.628 | 3.442 | 3.649 | 3.528 | 3.247 | 2.213 | 1.045 | 0.275 |
| 33 Pyridine ... Ethene | 0.699 | 1.487 | 1.764 | 1.762 | 1.621 | 1.013 | 0.367 | 0.047 |
| 34 Pentane ... Pentane | 2.789 | 3.569 | 3.734 | 3.581 | 3.278 | 2.223 | 1.050 | 0.273 |
| 35 Neopentane ... Pentane | 1.817 | 2.466 | 2.590 | 2.467 | 2.240 | 1.491 | 0.699 | 0.187 |
| 36 Neopentane ... Neopentane | 1.428 | 1.713 | 1.753 | 1.667 | 1.524 | 1.042 | 0.504 | 0.136 |
| 37 Cyclopentane ... Neopentane | 1.574 | 2.225 | 2.384 | 2.306 | 2.122 | 1.458 | 0.705 | 0.191 |
| 38 Cyclopentane ... Cyclopentane | 2.199 | 2.811 | 2.971 | 2.839 | 2.578 | 1.705 | 0.791 | 0.207 |
| 39 Benzene ... Cyclopentane | 2.013 | 3.135 | 3.496 | 3.420 | 3.135 | 2.062 | 0.903 | 0.194 |
| 40 Benzene ... Neopentane | 1.753 | 2.597 | 2.839 | 2.767 | 2.544 | 1.703 | 0.773 | 0.191 |
| 41 Uracil ... Pentane | 3.732 | 4.624 | 4.792 | 4.569 | 4.064 | 2.451 | 0.986 | 0.220 |
| 42 Uracil ... Cyclopentane | 2.963 | 3.910 | 4.094 | 3.895 | 3.526 | 2.298 | 1.026 | 0.253 |
| 43 Uracil ... Neopentane | 2.807 | 3.560 | 3.678 | 3.478 | 3.142 | 2.055 | 0.932 | 0.235 |
| 44 Ethene ... Pentane | 1.589 | 1.918 | 1.948 | 1.830 | 1.648 | 1.078 | 0.490 | 0.121 |
| 45 Ethyne ... Pentane | 0.997 | 1.526 | 1.669 | 1.616 | 1.473 | 0.958 | 0.418 | 0.097 |
| 46 Peptide ... Pentane | 3.673 | 4.144 | 4.195 | 4.009 | 3.700 | 2.615 | 1.190 | 0.290 |
| 47 Benzene ... Benzene (TS) | 1.549 | 2.513 | 2.823 | 2.790 | 2.590 | 1.772 | 0.839 | 0.230 |
| 48 Pyridine ... Pyridine (TS) | 2.444 | 3.248 | 3.492 | 3.426 | 3.202 | 2.294 | 1.188 | 0.378 |
| 49 Benzene ... Pyridine (TS) | 1.992 | 2.975 | 3.280 | 3.229 | 3.004 | 2.103 | 1.060 | 0.339 |
| 50 Benzene ... Ethyne (CH-π) | 1.784 | 2.585 | 2.822 | 2.764 | 2.563 | 1.785 | 0.895 | 0.274 |
| 51 Ethyne ... Ethyne (TS) | 1.183 | 1.449 | 1.506 | 1.449 | 1.335 | 0.930 | 0.462 | 0.135 |
| 52 Benzene ... AcOH (OH-π) | 3.895 | 4.523 | 4.661 | 4.519 | 4.228 | 3.121 | 1.705 | 0.558 |
| 53 Benzene ... AcNH$_2$ (NH-π) | 3.763 | 4.247 | 4.346 | 4.216 | 3.961 | 2.974 | 1.651 | 0.486 |
| 54 Benzene ... Water (OH-π) | 2.710 | 3.139 | 3.212 | 3.089 | 2.869 | 2.090 | 1.153 | 0.417 |
| 55 Benzene ... MeOH (OH-π) | 3.316 | 3.941 | 4.106 | 4.005 | 3.760 | 2.786 | 1.529 | 0.521 |
| 56 Benzene ... MeNH$_2$ (NH-π) | 2.376 | 2.983 | 3.153 | 3.062 | 2.823 | 1.939 | 0.941 | 0.264 |
| 57 Benzene ... Peptide (NH-π) | 3.620 | 4.866 | 5.220 | 5.105 | 4.759 | 3.419 | 1.818 | 0.626 |
| 58 Pyridine ... Pyridine (CH-N) | 2.890 | 3.887 | 4.194 | 3.921 | 3.474 | 2.199 | 1.025 | 0.281 |
| 59 Ethyne ... Water (CH-O) | 2.573 | 2.844 | 2.883 | 2.788 | 2.618 | 1.994 | 1.177 | 0.460 |
| 60 Ethyne ... AcOH (OH-π) | 4.295 | 4.788 | 4.863 | 4.692 | 4.385 | 3.248 | 1.769 | 0.557 |
| 61 Pentane ... AcOH | 2.642 | 2.852 | 2.839 | 2.697 | 2.488 | 1.775 | 0.784 | 0.171 |
| 62 Pentane ... AcNH$_2$ | 3.079 | 3.442 | 3.458 | 3.283 | 3.013 | 2.102 | 1.041 | 0.276 |
| 63 Benzene ... AcOH | 2.574 | 3.483 | 3.715 | 3.591 | 3.296 | 2.226 | 1.024 | 0.265 |
| 64 Peptide ... Ethene | 2.528 | 2.893 | 2.945 | 2.820 | 2.605 | 1.846 | 0.878 | 0.191 |
| 65 Pyridine ... Ethyne | 3.647 | 4.002 | 4.065 | 3.948 | 3.727 | 2.870 | 1.684 | 0.623 |
| 66 MeNH$_2$ ... Pyridine | 3.357 | 3.803 | 3.907 | 3.803 | 3.582 | 2.697 | 1.502 | 0.497 |